\documentclass[11pt]{article}
\usepackage{amsmath,amssymb,bm}
\usepackage{graphics}
\usepackage{graphicx}
\usepackage{amssymb}
\usepackage{epstopdf}

\usepackage{graphicx}
\usepackage{dcolumn}
\usepackage{bm}
\usepackage{amssymb}
\usepackage{xspace}

\usepackage{hyperref}
\usepackage{amsmath}

\def\pythia{{\sc Pythia}}

\usepackage{mystyle}
\usepackage{cite,./mcite}
\usepackage{authblk}

                \def\lsim{\mathrel{\rlap{\lower4pt\hbox{\hskip1pt$\sim$}}
    \raise1pt\hbox{$<$}}}                \def\gsim{\mathrel{\rlap{\lower4pt\hbox{\hskip1pt$\sim$}}
    \raise1pt\hbox{$>$}}}

\newcommand{\pt}{\ensuremath{p_{\rm{T}}}\xspace}

\title{Jet effects in high-multiplicity pp events}
\author[1]{Antonio Ortiz}
\author[1,2]{Gyula Benc\'edi}
\author[1,3]{H\'ector Bello}
\author[4]{Satyajit Jena}
{\tiny
\affil[1]{Instituto de Ciencias Nucleares, UNAM, 04510, M\'exico D. F., M\'exico}
\affil[2]{Wigner Research Centre for Physics of the H.A.S., H-1121, Budapest, Hungary}
\affil[3]{Facultad de Ciencias F\'isico Matem\'aticas, BUAP, 1152, Puebla, M\'exico}
\affil[4]{University of Houston, Houston, TX 77204, USA}
} 
\begin{document}

\maketitle 

\section{Introduction}

The study of the high-multiplicity pp events has become important because we need to understand the origin of the fluid-like features which have been found in such small systems~\cite{Ortiz:2015iha,Adam:2016dau,Armesto:2015ioy,Bautista:2015kwa}.

In this work we concentrate on the radial flow signatures,  which not only  hydrodynamical models can explain. Namely, the effect has been also found in  \pythia~\cite{Sjostrand:2014zea}  and  it is attributed to multi-parton interactions (MPI) and color reconnection (CR) via boosted color strings~\cite{Ortiz:2013yxa}. For high-multiplicity events, the blast-wave parametrization, a hydro inspired model, has been found to fit very well the transverse momentum (\pt) spectra of different particle species~\cite{Cuautle:2015kra}.  Although, the quality of the fits become worse for low-multiplicity events, we see that the parameter related to the average transverse expansion velocity ($\langle\beta_{\rm T}\rangle$) increases with increasing multiplicity. This effect is qualitatively similar to what has been seen at the LHC~\cite{Abelev:2013haa}.

In \pythia, color reconnection was originally introduced in order to explain the rise of the average \pt with the event multiplicity. In short, the model allows the interaction among the partons which originate from MPI and initial-/final-state radiation. There are different implementations, e.g.,  the default MPI-based model of \pythia 8.212 introduces a probability which is the largest for a  low-\pt system to be reconnected with one of a harder \pt scale.  And the interaction between two systems of high-\pt scales is not allowed.  Such a soft-hard interaction also suggests that jets may play a role in the observed radial flow-like patterns as highlighted in~\cite{Cuautle:2015kra,Ortiz:2015cma}.

In this work, the role of jets in high-multiplicity pp collisions is investigated using \pythia\,8.212. The inclusive \pt spectra of identified particles are studied for events with and without jets, where the jets are reconstructed using the anti-$k_{\rm T}$ algorithm implemented in FastJet~\cite{Cacciari:2011ma}.


\section{Results}

Proton-proton collisions at $\sqrt{s}=7$\,TeV were simulated with \pythia8.212 using the tune Monash 2013~\cite{Skands:2014pea}. Events were classified according with their event multiplicity ($N_{\rm ch}$) and leading jet \pt ($p_{\rm T}^{\rm jet}$). All the observables were calculated counting particles within $|\eta|$$<$1. For the jet finder only detectable particles (including charged and neutral particles) are considered within cone radius of 0.4, while for the \pt spectra and event multiplicity only charged particles are taken into account.

\begin{figure}[htbp]
\begin{center}
   \includegraphics[width=0.45\textwidth]{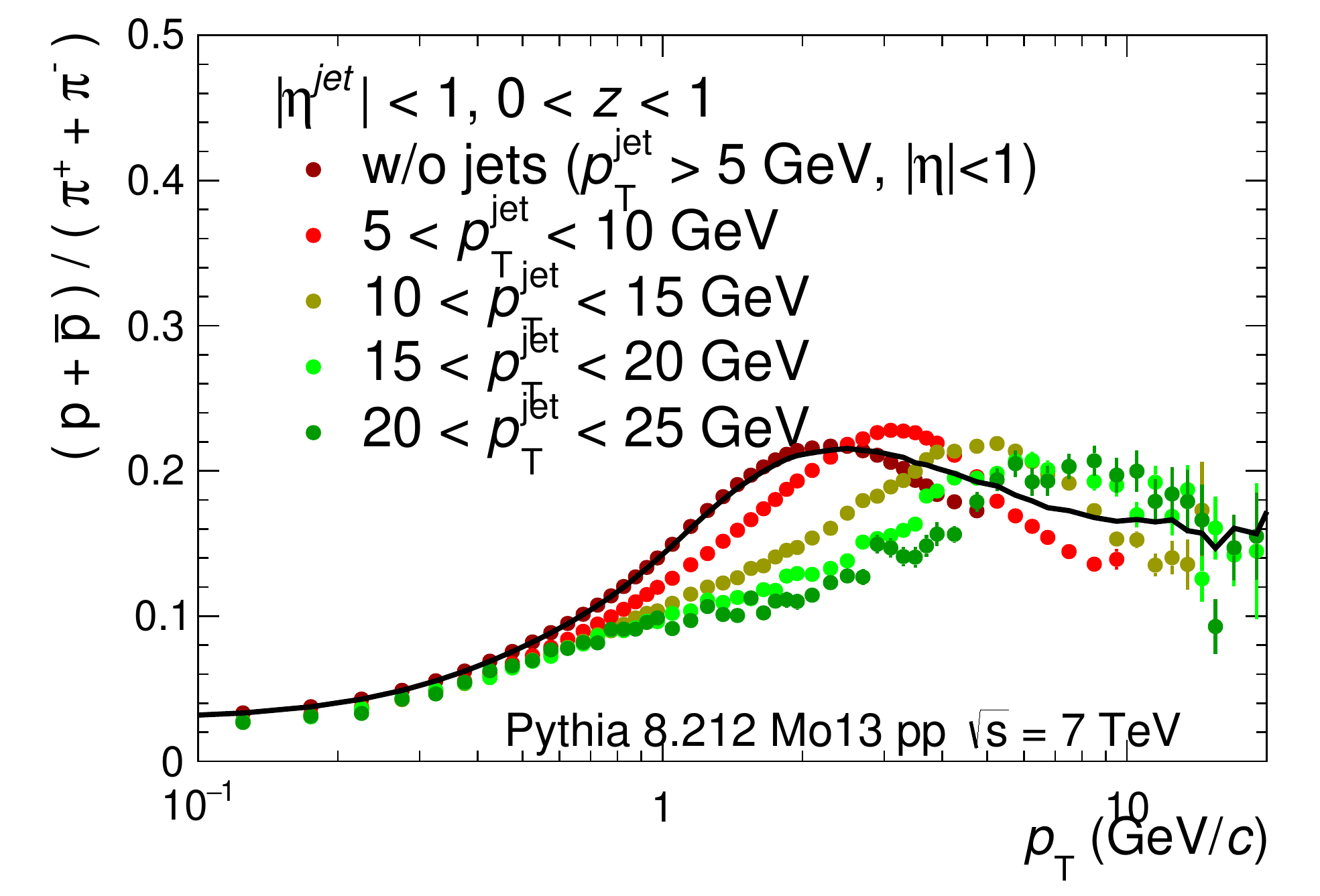}   
   \includegraphics[width=0.45\textwidth]{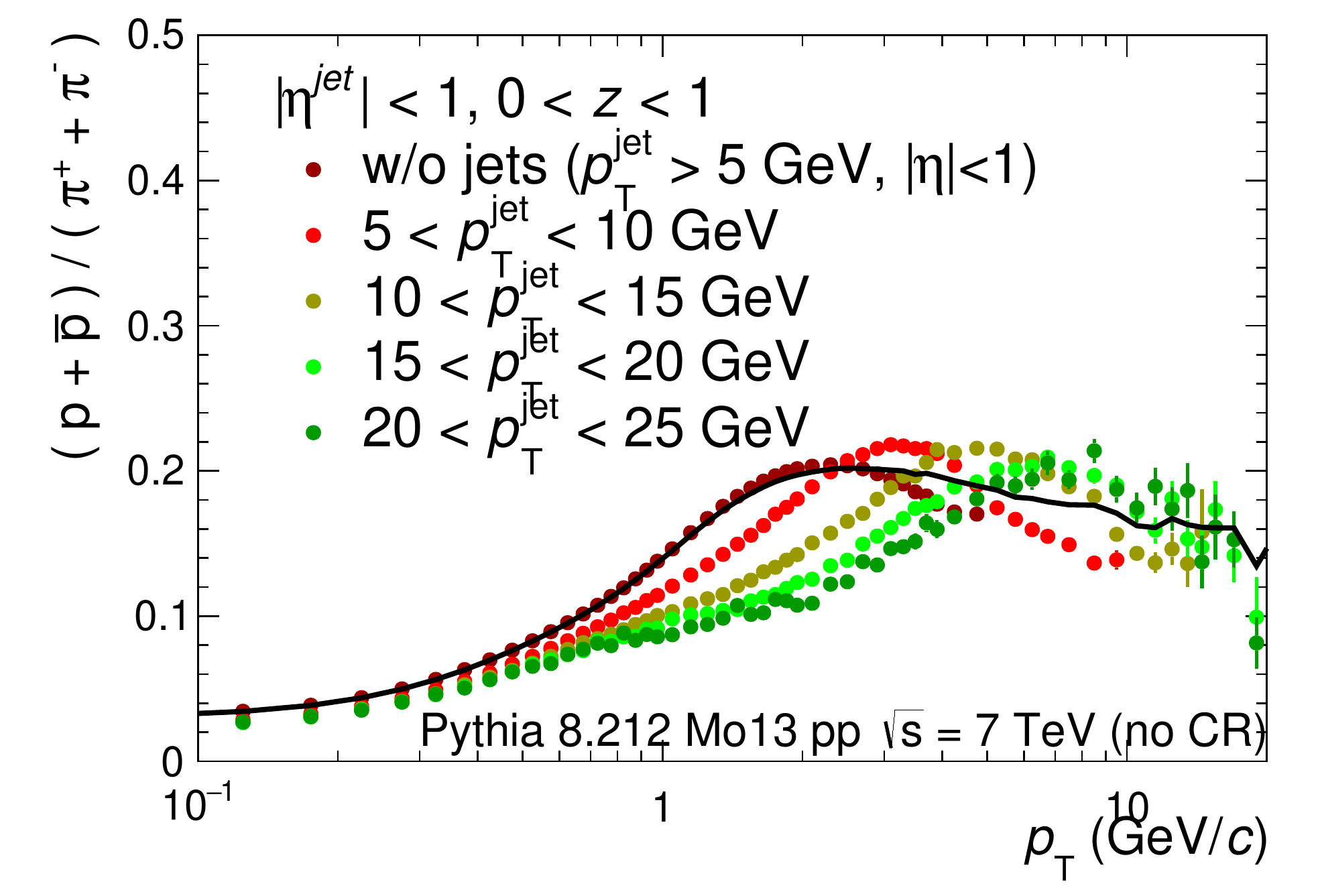}
   \caption{(Color online). Leading jet \pt dependence of the proton-to-pion ratio as a function of transverse momentum for low-multiplicity pp collisions at $\sqrt{s}=7$\,TeV. The results for the different $p_{\rm T}^{\rm jet}$ intervals (markers) are compared with the inclusive case (solid line). Events with (left) and without (right) color reconnection are shown.}
  \label{fig:1}
\end{center}
\end{figure}

To investigate on the radial flow-like effects in jets we first study the proton-to-pion ratio in low-multiplicity events and as a function of $p_{\rm T}^{\rm jet}$ (see Fig.~\ref{fig:1}). It is worth noticing that events without jets\footnote{Events without jets are those where the jet finder can not reconstruct one with $p_{\rm T}^{\rm jet}$$>$5\,GeV/$c$.} dominate for momenta below 2\,GeV/$c$, while at larger momenta, jets start playing a more important role. In addition, a bump at intermediate \pt is observed in all the event classes. The \pt, where the peak emerges, increases with increasing $p_{\rm T}^{\rm jet}$. This structure resembles one observed in the different colliding systems at the LHC~\cite{Adam:2015kca,Ortiz:2015iha} and which sometimes is referred as a ``flow peak''~\cite{Ortiz:2013yxa}. This effect is the same in events generated with and without color reconnection. 

\begin{figure}[htbp]
\begin{center}
   \includegraphics[width=0.32\textwidth]{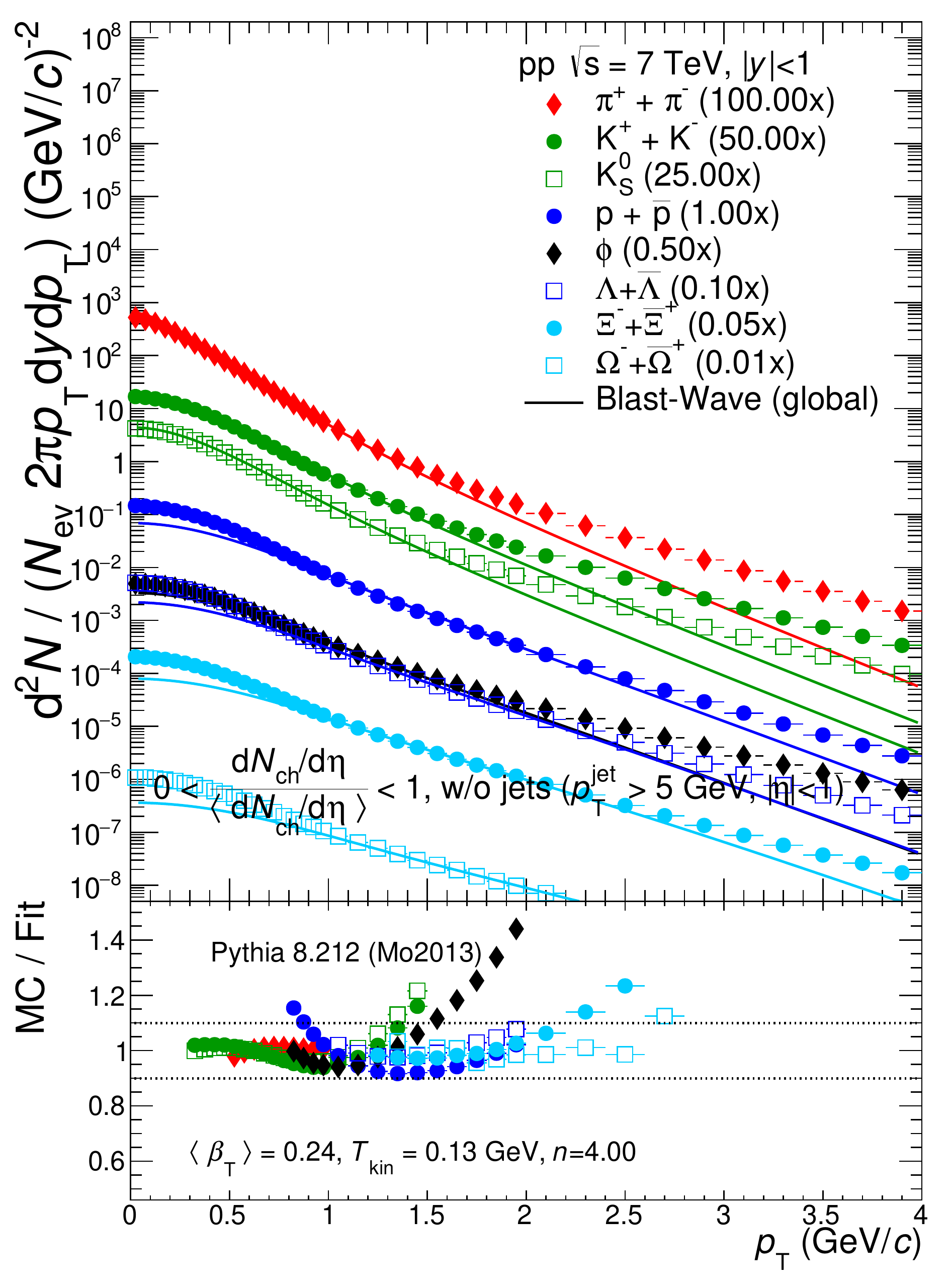}   
   \includegraphics[width=0.32\textwidth]{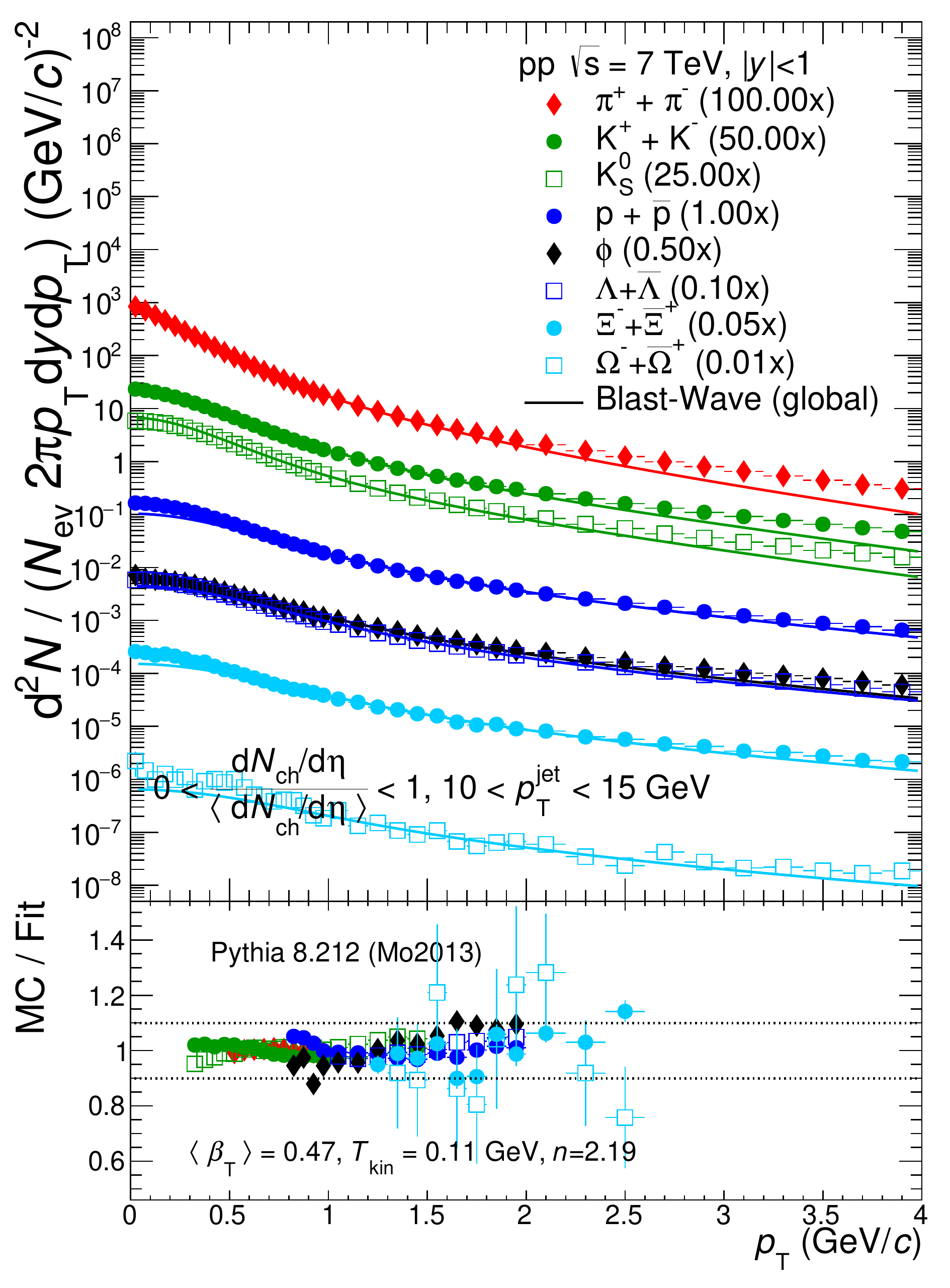}
   \includegraphics[width=0.32\textwidth]{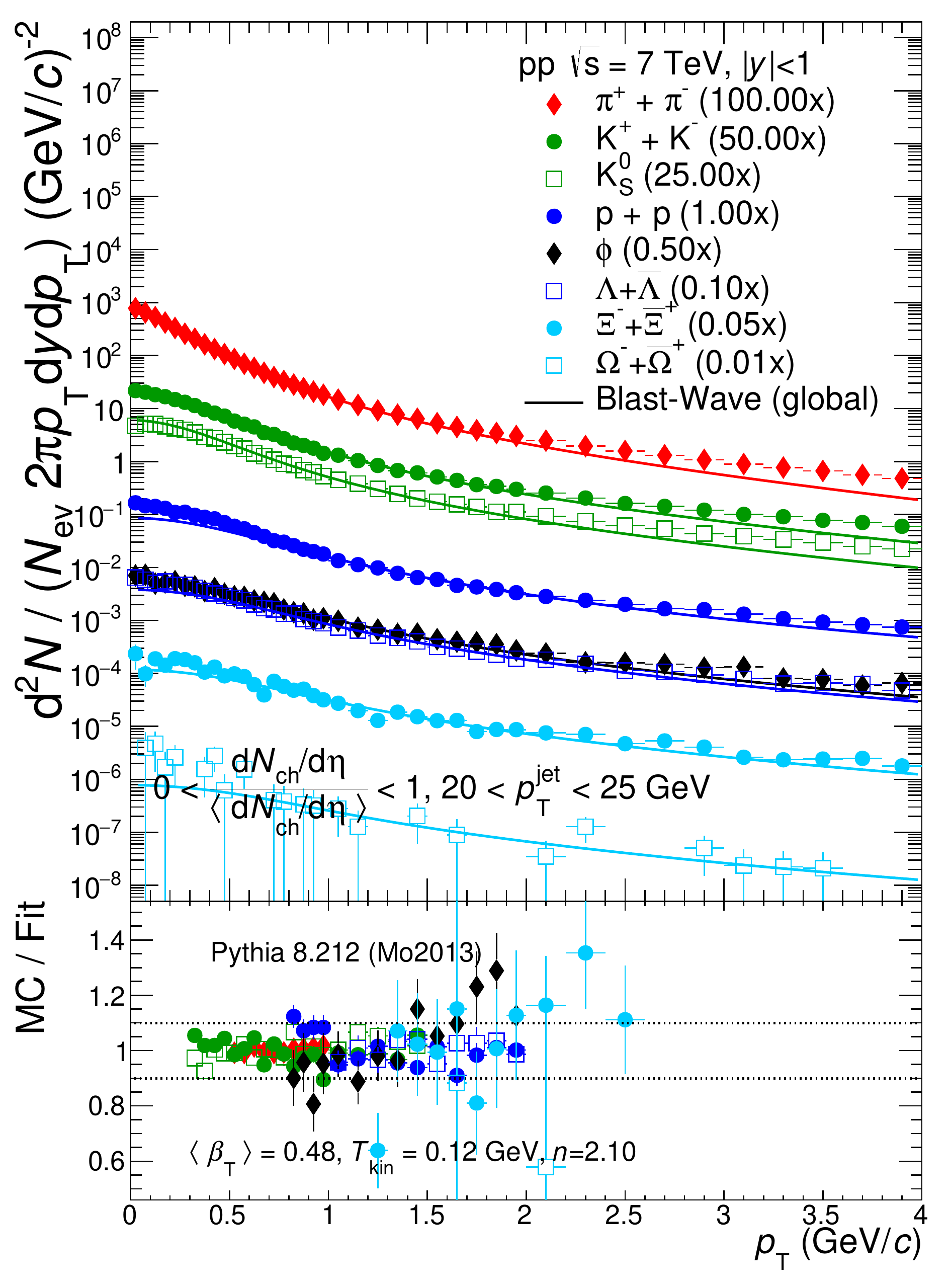}
   \includegraphics[width=0.32\textwidth]{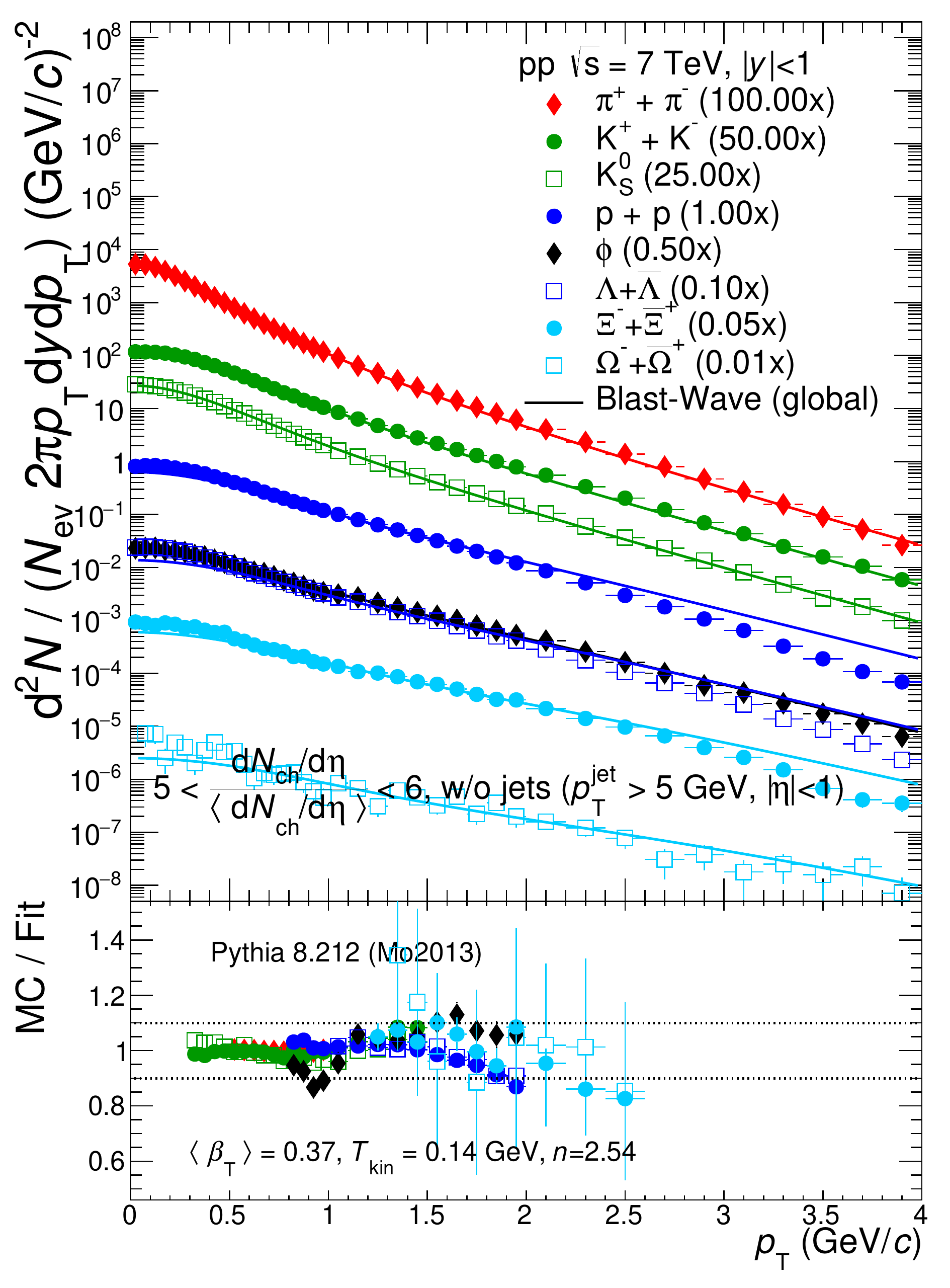}   
   \includegraphics[width=0.32\textwidth]{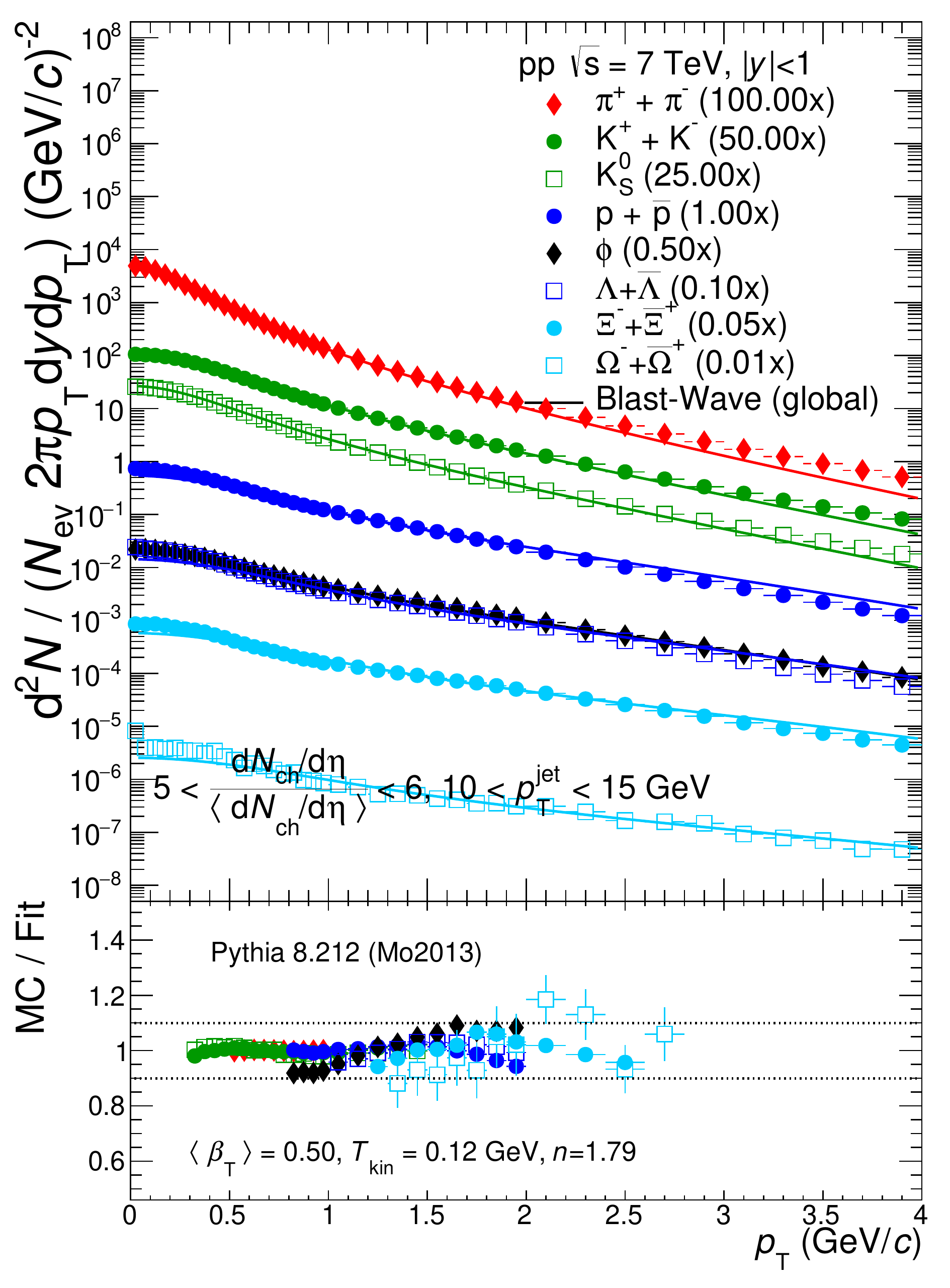}
   \includegraphics[width=0.32\textwidth]{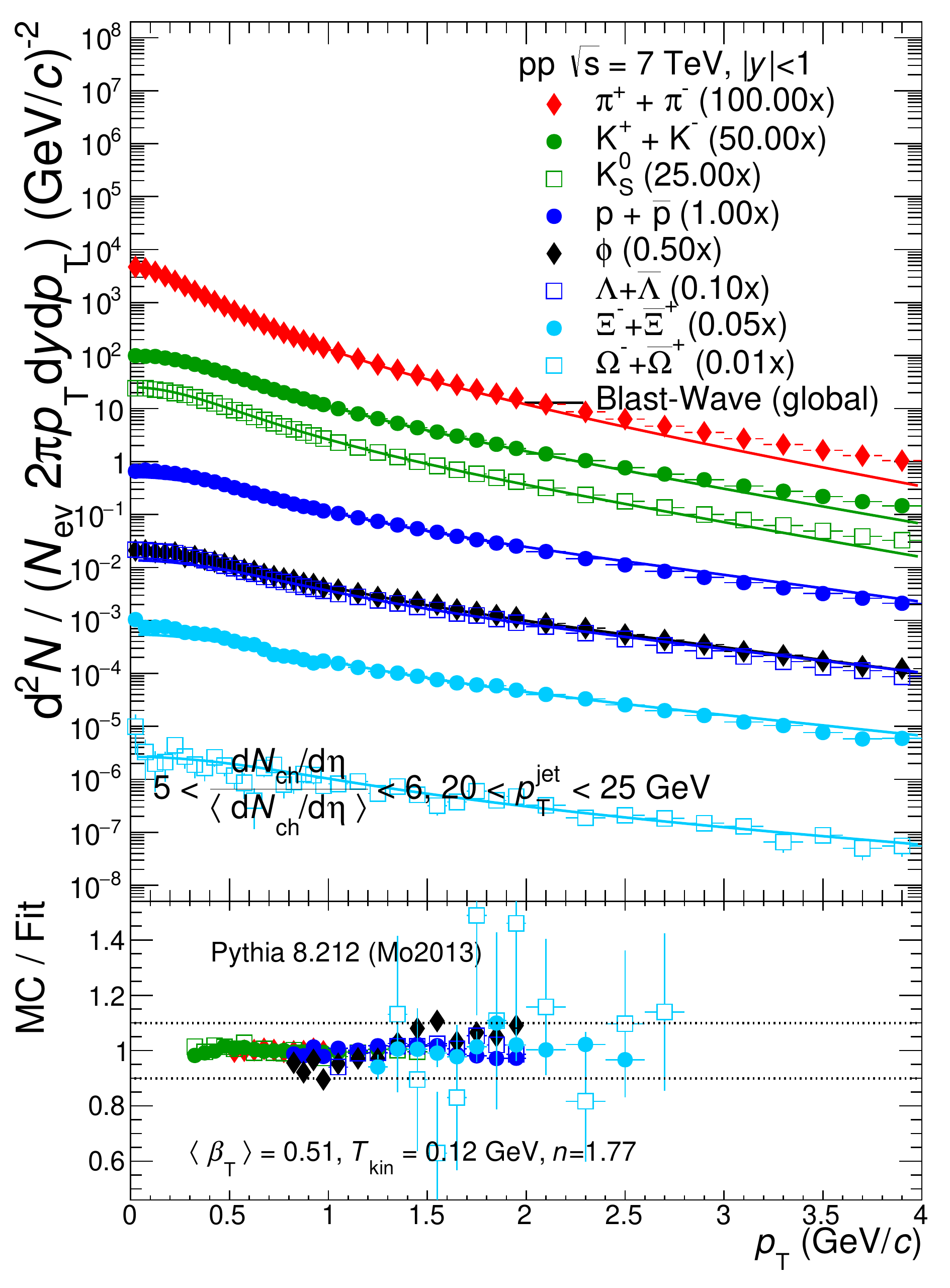}      
   \caption{(Color online).  Leading jet \pt dependence of the transverse momentum spectra for low (top) and high (bottom) multiplicity pp collisions at $\sqrt{s}=7$\,TeV. The blast-wave parametrization is shown with solid lines.}
  \label{fig:2}
\end{center}
\end{figure}

The blast-wave analysis of the \pt spectra has been performed using the same particle species and \pt intervals described in~\cite{Cuautle:2015kra}. Figure~\ref{fig:2} shows that the hydro model can describe the \pythia\, \pt spectra when jets with momentum above 5\,GeV/$c$ are part of the event. Actually, a $\langle \beta_{\rm T}\rangle$ of $\approx$0.5 can be achieved when the jet \pt is larger than 20\,GeV/$c$. Contrarily, the model does not describe the spectra in events without jets. This result is consistent with the spherocity analysis reported in~\cite{Cuautle:2015kra}, where it was argued that the fast parent parton being a boosted system can mimic radial flow too. The same analysis was also implemented for high-multiplicity events, in that case, thanks to color reconnection, the quality of the fit improves in events without jets, however a small $\langle \beta_{\rm T} \rangle$ ($\approx$0.37) is obtained and it increases up to $\approx$0.51 when a high-\pt jet is identified in the event. Actually, when a high-\pt jet was required, a very weak multiplicity dependence of $\langle \beta_{\rm T} \rangle$ is observed.

\section{Summary}

In summary, we have studied the role of jets in the radial flow-like features of \pythia. We have found that even in low-multiplicity events the blast-wave model is able to describe the \pt spectra of different particle species only when jets are part of the event. At high-multiplicity, $\langle \beta_{\rm T} \rangle$ can be very small in events without jets ($\approx$0.37). The interaction of jets with the soft component is therefore important to produce the observed effects in \pythia. This seems to be a promising tool which could be exploited by the experiments in order to understand better the LHC data.

\section*{Acknowledgements}
The authors acknowledge the useful discussions with Guy Pai\'c, Eleazar Cuautle, Peter Christiansen and Gergely Barnaf{\"o}ldi. Support  for  this  work  has  been  received  from  CONACYT  under  the  grant  No.   260440;  from
DGAPA-UNAM under PAPIIT grants IA102515, IN105113, IN107911 and IN108414; and OTKA under the grant NK106119. The EPLANET program supported the mobility from Mexico to Europe and vis. 

\bibliographystyle{MPI2015}
\bibliography{references}

\begin{thebibliography}{10}
\newcommand{\enquote}[1]{``#1''}
\providecommand{\url}[1]{\texttt{#1}}
\providecommand{\urlprefix}{URL }
\providecommand{\eprint}[2][]{\url{#2}}

\bibitem{Ortiz:2015iha}
A.~Ortiz (ALICE), \enquote{Overview of ALICE results,} \emph{Nuclear and
  Particle Physics Proceedings}, \textbf{267-269}(2015), 403, x Latin American
  Symposium of High Energy Physics.

\bibitem{Adam:2016dau}
Jaroslav Adam \emph{et~al.} (ALICE), \enquote{{Multiplicity dependence of
  charged pion, kaon, and (anti)proton production at large transverse momentum
  in p-Pb collisions at $\mathbf{\sqrt{{\textit s}_{\rm NN}}}$ = 5.02 TeV},}
  (2016), \eprint{1601.03658}.

\bibitem{Armesto:2015ioy}
N.~Armesto and E.~Scomparin, \enquote{{Heavy-ion collisions at the Large Hadron
  Collider: a review of the results from Run 1},} (2015), \eprint{1511.02151}.

\bibitem{Bautista:2015kwa}
I.~Bautista, A.~Fernandez, and P.~Ghosh, \enquote{{Indication of change of
  phase in high-multiplicity proton-proton events at LHC in String Percolation
  Model},} \emph{Phys. Rev.}, \textbf{D92}(2015)~(7), 071504.

\bibitem{Sjostrand:2014zea}
{T. Sj{\"o}strand, S. Ask, J. R. Christiansen, R. Corke, N. Desai, P. Ilten, S.
  Mrenna, S. Prestel, C. O. Rasmussen and P. Z. Skands}, \enquote{An
  Introduction to PYTHIA 8.2,} \emph{Comput. Phys. Commun.},
  \textbf{191}(2015), 159.

\bibitem{Ortiz:2013yxa}
{A. Ortiz, P. Christiansen, E. Cuautle, I. Maldonado, Ivonne and G. Pai{\'c}},
  \enquote{{Color Reconnection and Flowlike Patterns in pp Collisions},}
  \emph{Phys. Rev. Lett.}, \textbf{111}(2013)~(4), 042001.

\bibitem{Cuautle:2015kra}
{A. Ortiz, E. Cuautle and G. Pai{\'c}}, \enquote{{Mid-rapidity charged hadron
  transverse spherocity in pp collisions simulated with Pythia},} \emph{Nucl.
  Phys.}, \textbf{A941}(2015), 78.

\bibitem{Abelev:2013haa}
B.~Abelev et~al. (ALICE), \enquote{{Multiplicity Dependence of Pion, Kaon,
  Proton and Lambda Production in p-Pb Collisions at $\sqrt{s_{NN}}$ = 5.02
  TeV},} \emph{Phys. Lett.}, \textbf{B728}(2014), 25.

\bibitem{Ortiz:2015cma}
A.~Ortiz, \enquote{{Mean $p_{\rm T}$ scaling with $m/n_q$ at the LHC: Absence
  of (hydro) flow in small systems?}} \emph{Nucl. Phys.}, \textbf{A943}(2015),
  9.

\bibitem{Cacciari:2011ma}
G.~P.~Salam M.~Cacciari and G.~Soyez, \enquote{{FastJet User Manual},}
  \emph{Eur. Phys. J.}, \textbf{C72}(2012), 1896.

\bibitem{Skands:2014pea}
S.~Carrazza P.~Skands and J.~Rojo, \enquote{{Tuning PYTHIA 8.1: the Monash 2013
  Tune},} \emph{Eur. Phys. J.}, \textbf{C74}(2014)~(8), 3024.

\bibitem{Adam:2015kca}
J.~Adam et~al. (ALICE), \enquote{{Centrality dependence of the nuclear
  modification factor of charged pions, kaons, and protons in Pb-Pb collisions
  at $\sqrt{s_{\rm NN}}=2.76$ TeV},} (2015), \eprint{1506.07287}.

\end{thebibliography}

\end{document}